\documentclass{article}%
\usepackage{graphicx}
\usepackage{epsfig}
\usepackage{amsmath, amstext, amsfonts}
\usepackage{amssymb}
\usepackage{multicol}
\usepackage[latin1]{inputenc}
\usepackage{amsmath}
\usepackage{amsfonts}%
\begin{document}

\title{Periodic Sturm-Liouville problems related to two Riccati equations of constant coefficients}
\author{K.V. Khmelnytskaya, H.C. Rosu, and A. González\\Potosinian Institute of Science and Technology,\\Apdo Postal 3-74 Tangamanga, 78231 San Luis Potosí, Mexico}
\date{}
\maketitle

\begin{abstract}
{\footnotesize \ \noindent We consider two closely related Riccati equations
of constant parameters whose particular solutions are used to construct the
corresponding class of supersymmetrically-coupled second-order
differential equations. We solve analytically these parametric periodic
problems along the positive real axis. Next, the analytically solved model is
used as a case study for a powerful numerical approach that is employed here
for the first time in the investigation of the energy band structure of
periodic not necessarily regular potentials. The approach is based on the
well-known self-matching procedure of James (1949) and implements the spectral
parameter power series solutions introduced by Kravchenko (2008). We obtain
additionally an efficient series representation of the Hill discriminant based
on Kravchenko's series.
\newline}

{\footnotesize \noindent PACS numbers: 02.30Jr, 02.30.Hq, 11.30.Pb\newline}
{\footnotesize \noindent Keywords: Riccati equation; Bloch solutions; Hill discriminant \newline}


\end{abstract}

\section{Introduction}

\medskip

Natural periodic and quasiperiodic structures have drawn the attention of
mankind since old times \cite{self-made}. Nowadays, in the technological
world, the scientists, especially those involved in the area of material
science, bring forth at a tremendous pace new artificial specimens in which
finite periodic structures are the main components. A rich
mathematical background related to periodicity has been developed along the
years \cite{Eastham, Magnus}. The construction of new periodic potentials and
the analysis of their specific properties could be valuable in guiding the
modern technological design. Within this context the relationship between
Riccati equations and Sturm-Liouville problems has been firmly known at least
since more than half a century \cite{IH} and more recently led to
supersymmetric (SUSY) quantum mechanics \cite{Cooper,DF,Correa}.

In the present work we will discuss in some detail the periodic
Sturm-Liouville (PSL) problems generated by particular solutions of the very simple Riccati equations
of constant coefficients \cite{const-R}
\begin{equation}
\frac{dR}{dx}+R^{2}+k_{0}^{2}=0\label{ricc}%
\end{equation}
and
\begin{equation}
\frac{d\Phi}{dx}-2S\Phi +\Phi^{2}+k_{0}^{2}+S^{2}=0~,\label{ricc-S}%
\end{equation}
which is closely related to the first one. We obtain the exact PSL solutions
in the Bloch form. In general, when available, the analytic solutions of a PSL
equation expressed as Bloch functions allow one to analyze the band structure
of the problem. Of course, this is not always possible. In many cases, even if
the exact solutions are known their Bloch form is hard to disentangle. As a
consequence, other well-established approaches are frequently used in order to analyze the band
structure of the spectrum: either by means of the Hill discriminant (or
Lyapunov function) \cite{Magnus} or using the band structure parameter
introduced by James \cite{J49}. In general, for this goal, two linearly
independent solutions are required for all values of the spectral parameter.
Hence, it is essential to have an as simple form as possible for the sought
solutions in terms of the spectral parameter. Therefore we show that a
convenient study of the band structure of the spectrum can be achieved using a
representation for solutions in the form of power series with respect to the
spectral parameter.

The outline of the paper is the following. The next section is devoted to the
three main points: (i) the SUSY construction of the periodic Sturm-Liouville
problems of Schr\"odinger type starting from the aforementioned Riccati
equation, (ii) solving the resulting Schr\"odinger equation and (iii) studying
some of its properties. Next, in Section 3, we provide a numerical approach of
the same problems using the self-matching method of H.M. James \cite{J49} for
the Kravchenko representation of the solutions in terms of spectral parameter
power series (SPPS) \cite{vvk, KP}. A small Conclusion section ends up the paper.

\section{\bigskip\ Riccati-associated PSL equations}

\subsection{Case I: Riccati equation (\ref{ricc})}

In the Riccati equation (\ref{ricc}) we introduce $R(x)=\frac{f^{\prime}}{f} $
leading to the second order linear differential equation
\begin{equation}
-f^{\prime\prime}(x)+\nu_{1}f(x)=0~,\label{2-f}%
\end{equation}
where $\nu_{1}=-k_{0}^{2}$.

The linearly independent solutions of (\ref{2-f}) are obviously
\[
f_{1}(x)=\cos k_{0}x~,\qquad f_{2}(x)=\sin k_{0}x~.
\]
Using $f_{1}$ one gets for the Riccati equation the solution $R_{f_{1}%
}(x)=-k_{0}\tan k_{0}x$. In what follows we employ $f_{1}$ to define
$R_{f_{1}}(x)$, equally well one can use $f_{2}$ leading to $R_{f_{2}%
}(x)=k_{0}\cot k_{0}x$, however this brings in only minimal changes in the whole
of the mathematical apparatus that follows and hereby we will deal only with
the first choice.

Since $\nu_{1}$ can be calculated by employing the equation
\[
\nu_{1}=\frac{dR_{f_{1}}}{dx}+R_{f_{1}}^{2}=-k_{0}^{2}~,
\]
the supersymmetric partner potential will be given by
\[
\nu_{2}(x)=-\frac{dR_{f_{1}}}{dx}+R_{f_{1}}^{2}=k_{0}^{2}(1+2\mathrm{\tan}%
^{2}k_{0}x)
\]
%
and the supersymmetric partner equation of equation~(\ref{2-f}) is given by
\begin{equation}
-g^{\prime\prime}(x)+\nu_{2}(x)g(x)=0~.\label{2-g}%
\end{equation}
The supersymmetric partner linear independent solutions are
\[
g_{1}(x)=\frac{1}{\cos k_{0}x}~,\qquad g_{2}(x)=\frac{1}{k_{0}\cos k_{0}%
x}\left[  \frac{k_{0}x}{2}+\frac{1}{4}\sin2k_{0}x\right]  ~.
\]

Considering now the spectral issue for these two periodic potentials, the
constant one $\nu_{1}$ and the singular one $\nu_{2}(x)$:%
\begin{align*}
-f^{\prime\prime}(x)-k_{0}^{2}f(x) &  =K^{2}f(x),\\
-g^{\prime\prime}(x)+k_{0}^{2}(1+2\mathrm{\tan}^{2}k_{0}x)g(x) &  =K^{2}g(x),
\end{align*}
we can get the $K^{2}$ spectrum of the $f$ problem from the Bloch solutions
$e^{\pm i\sqrt{k_{0}^{2}+K^{2}}x}$ that provide the quasimomentum
$P(K^{2})=\sqrt{k_{0}^{2}+K^{2}}$ .\medskip\ On the other hand, for the $g$
problem, being the Darboux partner of the $f$ problem, i.e., $g(x)=(\frac
{d}{dx}-R_{f_{1}}(x))f(x)$, we have the following Bloch solutions with the
same quasimomentum $P(K^{2})$
\[
e^{\pm i\sqrt{k_{0}^{2}+K^{2}}x}\left(  \mathrm{\tan}k_{0}x\pm i\sqrt
{k_{0}^{2}+K^{2}}\right)  .
\]
As known \cite{Shubin}, the allowed energy bands exist only for $P\in
\mathbf{R}$ leading to $K^{2}\geq-k_{0}^{2}$, therefore there is only one
forbidden zone covering the interval $(-\infty,-k_{0}^{2})$.

\medskip

\noindent Using the following two Pauli matrices
\[
\sigma_{y}=\left(
\begin{array}
[c]{cc}%
0 & -i\\
i & 0
\end{array}
\right)  \qquad\mathrm{and}\quad\sigma_{x}=\left(
\begin{array}
[c]{cc}%
0 & 1\\
1 & 0
\end{array}
\right)  ~,
\]
a single matrix equation for the two-component spinor $W=\left(  \mathrm{%
\begin{array}
[c]{c}%
g_{1}\\
f_{1}%
\end{array}
}\right)  $ can be written down:
\begin{equation}
\sigma_{y}W^{\prime}(x)+i\sigma_{x}R_{f_{1}}W(x)=0~.\label{HD}%
\end{equation}
The two components of the spinor $W$ have a sort of hidden coupling through the same particular Riccati solution but otherwise they look independent
one of the other.

\subsection{Case II: Riccati equation (\ref{ricc-S})}

\noindent A simple way to perform a direct coupling of the two components $f$ and $g$ is to add two
constant parameters as follows: $K$ as a spectral parameter of the Dirac-like
equations and a potential coupling parameter $S$
\begin{equation}
\lbrack\sigma_{y}\frac{d}{dx}+i\sigma_{x}(R_{f_{1}}+S)]W_{c}(x)=iKW_{c}%
(x)~,\qquad W_{c}=\left(  \mathrm{%
\begin{array}
[c]{c}%
g_{c}\\
f_{c}%
\end{array}
}\right)  .\label{HDM}%
\end{equation}
For related mathematical procedures the reader is directed to a textbook of
Lanczos \cite{L-LDOs} and to a paper of Nogami and Toyama \cite{NogToy93} and references therein
for similar supersymmetric structure of the Dirac equation in particle physics. Equation~(\ref{HDM}) is equivalent to the following
system of coupled equations
\begin{align}
g_{c}^{\prime}(x)+(k_{0}R_{f_{1}}+S)g_{c}(x) &  =Kf_{c}(x)\label{DT1}\\
f_{c}^{\prime}(x)-(k_{0}R_{f_{1}}+S)f_{c}(x) &  =-Kg_{c}(x).\label{DT2}%
\end{align}
This leads to:
\begin{align}
-\left(\frac{d}{dx}+\Phi_{1}\right)\left(\frac{d}{dx}-\Phi_{1}\right)f_{c} &  =K^{2}f_{c}%
(x),\label{f}\\
-\left(\frac{d}{dx}-\Phi_{1}\right)\left(\frac{d}{dx}+\Phi_{1}\right)g_{c} &  =K^{2}g_{c}%
(x),\label{g}%
\end{align}
where $\Phi_{1}=R_{f_{1}}+S=-k_{0}\tan k_{0}x+S$ is a particular solution of
the Riccati equation (\ref{ricc-S}). In unfactorized form, (\ref{f}) and
(\ref{g}) turn into the following equations
\begin{align}
-f_{c}^{\prime\prime}(x)+(-k_{0}^{2}+S^{2}-2Sk_{0}\tan k_{0}x)f_{c}(x) &
=K^{2}f_{c}(x)\label{S-2comp}\\
-g_{c}^{\prime\prime}(x)+(k_{0}^{2}+S^{2}-2Sk_{0}\tan k_{0}x+2k_{0}^{2}%
\tan^{2}k_{0}x)g_{c}(x) &  =K^{2}g_{c}(x)~.\label{S-1comp}%
\end{align}

The latter two equations define two new classes of parametric singular
potentials. To see the changes with respect to the initial equations
(\ref{2-f}) and (\ref{2-g}) for $f$ and $g$, respectively, we write the
previous system in terms of modified $\nu$ functions $\nu_{1c}(x)$ and
$\nu_{2c}(x) $:
\begin{equation}
-f_{c}^{\prime\prime}(x)+\nu_{1,c}(x)f_{c}(x)=K^{2}f_{c}(x),\qquad
-g_{c}^{\prime\prime}(x)+\nu_{2,c}(x)g_{c}(x)=K^{2}g_{c}(x)~,\label{nu's}%
\end{equation}
where
\[
\nu_{1,c}(x)=\nu_{1}+\Delta\nu(x),\qquad\nu_{2,c}(x)=\nu_{2}(x)+\Delta
\nu(x),\qquad\Delta\nu(x)=S^{2}-2Sk_{0}\tan k_{0}x~.
\]

Note that all the solutions $f_{c}$ of (\ref{S-2comp}) for any value of
$K^{2}$ are square integrable on any finite interval. This is due to Weyl's
alternative theorem, see for example the book of Hellwig \cite{hellwig}, all
singular points $\frac{n\pi}{2k_{0}},$ ($n$ odd)\ of $\nu_{1,c}(x)$ are limit
circle points. Indeed, it is sufficient to prove square integrability of two
linearly independent solutions for any fixed value of $K^{2}$. Taking
$K^{2}=0$ we see that both linearly independent solutions%
\[
h_{1}(x)=e^{Sx}\cos k_{0}x\quad\text{and}\quad h_{2}(x)=e^{Sx}\cos k_{0}%
x\int\frac{e^{-2Sx}}{\cos^{2}k_{0}x}dx\text{ }%
\]
are square integrable on any finite interval.

The limit circle points of equation (\ref{S-2comp}) become limit point
singularities of (\ref{S-1comp}). This can be demonstrated considering the
solution $\frac{1}{h_{1}(x)}$ of (\ref{S-1comp}) for $K^{2}=0$. Thus, Weyl's
alternative guarantees that exactly one square integrable solution of
(\ref{S-1comp})\ exists for $K^{2}$ with a nonvanishing imaginary part whereas
for real $K^{2}$ such a solution cannot even exist.

In order to solve the equations (\ref{S-2comp}),(\ref{S-1comp}) we begin with the
first equation in (\ref{nu's}). With the aid of its solution $f_{c}$ the
solution of the second equation in (\ref{nu's}) can be simply obtained by
applying the Darboux transformation
\begin{equation}
g_{c}(x)=f_{c}^{\prime}(x)-\Phi_{1}(x)f_{c}(x).\label{trans g}%
\end{equation}
Thus, we focus next on the $f_{c}$ equation which will be reduced to a
hypergeometric equation. The reduction is done with the help of a procedure
similar to the one described in \cite{Pert}.

\subsection{Hypergeometric solutions}

Consider the simultaneous change of the independent variable $\chi=\frac{1}%
{2}\left(  1-i\tan k_{0}x\right)  =\frac{e^{-ik_{0}x}}{2\cos k_{0}x}$ and of
the dependent variable $y(\chi)=(\chi^{2}-\chi)^{1/2}f_{c}$. Then the first
equation in (\ref{nu's}) takes the following Schrödinger-like
form
\begin{equation}
\frac{d^{2}y}{d\chi^{2}}+I_{f_{c}}y=0\label{hyp-schr}%
\end{equation}
%
where
\[
I_{f_{c}}=\frac{S^{2}-K^{2}+2iSk_{0}-4iSk_{0}\chi}{4k_{0}^{2}\chi^{2}%
(\chi-1)^{2}}~.
\]
Finally, in order to bring (\ref{hyp-schr}) to the hypergeometric form, the
following substitution can be used \cite{Kamke}
\[
y(\chi)=\chi^{p}(\chi-1)^{q}U(\chi)~,
\]
where
\[
p_{1,2}=\frac{1}{2}\left(  1\pm\sqrt{\frac{(k_{0}-iS)^{2}+K^{2}}{k_{0}^{2}}%
}\right)  ,\quad q_{1,2}=\frac{1}{2}\left(  1\pm\sqrt{\frac{(k_{0}%
+iS)^{2}+K^{2}}{k_{0}^{2}}}\right)  ~.
\]
Thus, one gets
\begin{equation}
\chi(\chi-1)\frac{d^{2}U}{d\chi^{2}}+2[(p_{1,2}+q_{1,2})\chi-p_{1,2}]\frac
{dU}{d\chi}+\left[  2p_{1,2}q_{1,2}-\frac{S^{2}-K^{2}}{2k_{0}}\right]
U=0\label{hyp-schr1}%
\end{equation}
%
Choosing the pair $p_{1}$ and $q_{1}$, we obtain the following solutions
\begin{equation}
U_{1}={}_{2}F_{1}(p_{1}+q_{1}-1,p_{1}+q_{1};2p_{1};\chi)~,\qquad
U_{2}=x^{1-2p_{1}}{}_{2}F_{1}(-p_{1}+q_{1}+1,-p_{1}+q_{1};2-2p_{1}%
;\chi)~.\label{2-sol1}%
\end{equation}
Using properties of hypergeometric functions (the change of variable
$\chi\rightarrow\frac{\chi}{\chi-1}$), the $f_{c}$ linearly independent
solutions can be written in the form
\begin{equation}
f_{c,1}(x)=(-1)^{q_{1}-\frac{1}{2}}e^{-i(2p_{1}-1)k_{0}x}\,{}_{2}F_{1}\left(
p_{1}+q_{1}-1,p_{1}-q_{1};2p_{1};-e^{-2ik_{0}x}\right)  ~,\label{2-fcfin1}%
\end{equation}
%
\begin{equation}
f_{c,2}(x)=(-1)^{q_{1}-\frac{1}{2}}e^{i(2p_{1}-1)k_{0}x}\,{}_{2}F_{1}\left(
-p_{1}+q_{1},-p_{1}-q_{1}+1;2-2p_{1};-e^{-2ik_{0}x}\right)  ~.\label{2-fcfin2}%
\end{equation}

Application of the Darboux transformation (\ref{trans g}) \ to the solutions
$f_{c,1}(x)$ and $f_{c,2}(x)$\ leads to the following solutions $g_{c,1}(x)$
and $g_{c,2}(x)$ of the second equation in (\ref{nu's})%

\begin{align}
g_{c,1}(x) &  =(-1)^{q_{1}-\frac{1}{2}}e^{-i(2p_{1}-1)k_{0}x}\,\times
\nonumber\\
&  \times{}\left[  (k_{0}\tan k_{0}x-S-ik_{0}(2p_{1}-1)\,)\,_{2}F_{1}\left(
p_{1}+q_{1}-1,\,p_{1}-q_{1};2p_{1};-e^{-2ik_{0}x}\right)  ~+\right.
\nonumber\\
&  \left.  +\frac{(p_{1}+q_{1}-1)(p_{1}-q_{1})}{p_{1}}ik_{0}e^{-2ik_{0}%
x}\,_{2}F_{1}\left(  p_{1}+q_{1},\,p_{1}-q_{1}+1;\,2p_{1}+1;\,-e^{-2ik_{0}%
x}\right)  \right]  ,\label{g1}%
\end{align}

\begin{align}
g_{c,2}(x) &  =(-1)^{q_{1}-\frac{1}{2}}e^{i(2p_{1}-1)k_{0}x}\times\nonumber\\
&  \times\,{}\left[  (k_{0}\tan k_{0}x-S+ik_{0}(2p_{1}-1)\,)\,_{2}F_{1}\left(
-p_{1}+q_{1},-p_{1}-q_{1}+1;2-2p_{1};-e^{-2ik_{0}x}\right)  ~+\right.
\nonumber\\
&  \left.  +\frac{(p_{1}+q_{1}-1)(p_{1}-q_{1})}{1-p_{1}}ik_{0}e^{-2ik_{0}%
x}\,_{2}F_{1}\left(  -p_{1}+q_{1}+1,\,-p_{1}-q_{1}+2;\,3-2p_{1};\,-e^{-2ik_{0}%
x}\right)  \right]  ,\label{g2}%
\end{align}
Though the Darboux transformation is applied to regular solutions $f_{c,1}$
and $f_{c,2}$ the singularity of the superpotential $\Phi$ implies the
singularity of $g_{c,1}$ and $g_{c,2}$. \bigskip

The solutions (\ref{2-fcfin1}), (\ref{2-fcfin2}) and (\ref{g1}), (\ref{g2})
are quasiperiodic or Bloch functions with the quasimomentum
\[
P_{S}(K^{2})=(2p_{1}-1)k_{0}=\sqrt{(k_{0}-iS)^{2}+K^{2}},
\]
which defines the Brillouin zone as follows: $\operatorname{Re}(P_{S}%
)\in\lbrack-k_{0},k_{0}]$ . The allowed energies exist only for $P_{S}%
\in\mathbf{R}$ \cite{Shubin}. This condition holds when $(k_{0}-iS)^{2}%
+K^{2}\in\mathbf{R}^{+}$. To specify the spectrum we should make some
additional considerations. Limiting ourselves to the real values of the
spectral parameter $K^{2}$ we have that for $k_{0}\in\mathbf{R}$, $S$ should
be of the form $S=is$, where $s\in\mathbf{R}$. Moreover, $K^{2}$ must satisfy
the inequality $K^{2}\geq-(k_{0}+s)^{2}$ or equivalently $K^{2}\in
(-(k_{0}+s)^{2},\infty]$. It is worth mentioning that for $s=0$ we get the
spectrum of the uncoupled potentials $\nu_{1}$ and $\nu_{2}$. Another
apparently possible case: $S\in\mathbf{R}$, $k_{0}=i\gamma$, where $\gamma
\in\mathbf{R}$ that also leads to a real $P_{S}$ is not meaningful for the
analysis of issues related to the quasimomentum since the potentials
$\nu_{1,c}$ and $\nu_{2,c}$\ are not periodic any more. Thus the spectrum of
(\ref{S-2comp}) and (\ref{S-1comp}) is real only for purely imaginary values
of the parameter $S$.

Notice that periodic potentials with purely imaginary coupling constants have
been considered for the case of Mathieu equation \cite{Mulholland} and have
applications to the alternating flow of electromagnetic fields along
conducting elliptic cylinders.

At the band edges we have $P_{S}=\left\{
\begin{array}
[c]{c}%
2nk_{0}\\
(2n+1)k_{0}%
\end{array}
\right.  ,n=0,\pm1,\pm2,\cdots$. The solutions $f_{c,1}$ and $f_{c,2}$ are
periodic when
\begin{equation}
P_{S}=2nk_{0}\Longrightarrow K^{2}=(2nk_{0})^{2}-(k_{0}-iS)^{2}\label{per}%
\end{equation}
\ and antiperiodic for
\begin{equation}
P_{S}=(2n+1)k_{0}\Longrightarrow K^{2}=((2n+1)k_{0})^{2}-(k_{0}-iS)^{2}%
.\label{antiper}%
\end{equation}
For these special values of $K^{2}$ corresponding to the band edges two pairs
of Bloch solutions $f_{c,1}(x,K^{2}),$ $f_{c,2}(x,K^{2})$ and $g_{c,1}%
(x,K^{2}),$ $g_{c,2}(x,K^{2})$\ degenerate to single solutions $f_{c,1}%
(x,K^{2})\equiv$ $f_{c,2}(x,K^{2})$ and $g_{c,1}(x,K^{2})\equiv$
$g_{c,2}(x,K^{2})$. Indeed, the solutions (\ref{2-fcfin1}) and (\ref{2-fcfin2}%
) in the periodic case $P_{S}=2nk_{0}$ have the form%
\[
f_{c,1}^{per}(x)=(-1)^{q_{1}(n)-\frac{1}{2}}e^{-2ink_{0}x}\,{}_{2}F_{1}\left(
n-\frac{1}{2}+q_{1}(n),n+\frac{1}{2}-q_{1}(n);1+2n;-e^{-2ik_{0}x}\right)  ~,
\]
%
\[
f_{c,2}^{per}(x)=(-1)^{q_{1}(m)-\frac{1}{2}}e^{2imk_{0}x}\,{}_{2}F_{1}\left(
-m-\frac{1}{2}+q_{1}(m),-m+\frac{1}{2}-q_{1}(m);1-2m;-e^{-2ik_{0}x}\right)  ~,
\]
where $q_{1}(n)=\frac{1}{2}+\sqrt{n^{2}+i\frac{S}{k_{0}}}$.

\bigskip It is clear that\ in order to have the same value of $P_{S}$ in both
functions we should \ take $m=-n$, thus the solutions are periodic and
$f_{c,1}^{per}(x)\equiv f_{c,2}^{per}(x)$.

In the antiperiodic case $P_{S}=(2n+1)k_{0}$ we have $q_{1}(n)=\frac{1}%
{2}+\sqrt{(n+\frac{1}{2})^{2}+i\frac{S}{k_{0}}}$ and%

\[
f_{c,1}^{aper}(x)=(-1)^{q_{1}(n)-\frac{1}{2}}e^{-i(2n+1)k_{0}x}\,{}_{2}%
F_{1}\left(  n+q_{1}(n),n+1-q_{1}(n);2n+2;-e^{-2ik_{0}x}\right)  ~,
\]
\[
f_{c,2}^{aper}(x)=(-1)^{q_{1}(m)-\frac{1}{2}}e^{i(2m+1)k_{0}x}\,{}_{2}%
F_{1}\left(  -m-1+q_{1}(m),-m-q_{1}(m);-2m;-e^{-2ik_{0}x}\right)  ~.
\]

Taking $m=-(n+1)$ we obtain that $f_{c,1}^{aper}\equiv f_{c,2}^{aper}$ are
antiperiodic. Note that in both cases $q_{1}(n)=q_{1}(m)$. Analogously we
obtain%
\begin{align*}
g_{c}^{per}(x) &  =\left.  (-1)^{q_{1}(n)-\frac{1}{2}}e^{-2ink_{0}x}\right[
\,(k_{0}\tan k_{0}x-S-2ink_{0})\ \times\\
&  \times_{2}F_{1}\left(  n-\frac{1}{2}+q_{1}(n),n+\frac{1}{2}-q_{1}%
(n);1+2n;-e^{-2ik_{0}x}\right)  +\\
&  \left.  +\frac{S}{n+\frac{1}{2}}{}e^{-2ik_{0}x}~_{2}F_{1}\left(  n+\frac
{1}{2}+q_{1}(n),n+\frac{3}{2}-q_{1}(n);2+2n;-e^{-2ik_{0}x}\right)  \right]  ,
\end{align*}
and%
\begin{align*}
g_{c}^{aper}(x) &  =\left.  (-1)^{q_{1}(n)-\frac{1}{2}}e^{-i(2n+1)k_{0}%
x}\right[  (k_{0}\tan k_{0}x-S-(2n+1)ik_{0})\,{}\times\\
&  \times_{2}F_{1}\left(  n+q_{1}(n),n+1-q_{1}(n);2n+2;-e^{-2ik_{0}x}\right)
+\\
&  \left.  +\frac{S}{n+1}{}e_{2}^{-2ik_{0}x}\ _{2}F_{1}\left(  n+1+q_{1}%
(n),n+2-q_{1}(n);2n+3;-e^{-2ik_{0}x}\right)  \right]  .
\end{align*}

The value $P_{S}=0$\ and consequently $K_{0}^{2}=-(k_{0}+s)^{2}$\ give us the
following periodic nodeless solution of (\ref{S-2comp}) which we denote by
$f_{0}(x)$
\begin{equation}
f_{0}(x)=(-1)^{i\sqrt{\frac{s}{k_{0}}}}\,{}_{2}F_{1}\left(  i\sqrt{\frac
{s}{k_{0}}},-i\sqrt{\frac{s}{k_{0}}};1;-e^{-2ik_{0}x}\right)  .\label{f0}%
\end{equation}
This eigenfunction will be used later on.

\section{An efficient numerical approach for the energy band structure}

The potentials $\nu_{1,c}$ and $\nu_{2,c}$\ are periodic functions of period
$\Lambda=\frac{\pi}{k_{0}}$, and have singularities at the points $\frac{n\pi
}{2k_{0}}$. Following James \cite{J49} we choose the first period as
$[0,\Lambda]$ and call it the zeroth cell, the second period $[\Lambda
,2\Lambda]$\ as the first cell, and so forth. Following the fundamental
procedure of James \cite{J49}, we construct the so-called self-matching
solutions of the SUSY-related equations (\ref{S-2comp}) and (\ref{S-1comp})
for the zeroth cell which allows us to build the Bloch solutions on the entire
range of $x$. We proceed further by writing the SPPS representation
of the associated Hill discriminants which allows us to describe the spectrum
of the SUSY-related equations (\ref{S-2comp}) and (\ref{S-1comp}) in a simple
way. For doing this we choose to use a numerically calculated solution instead
of the exact one given in the preceding section. First, because the SPPS
approach is clearly more universal and can be applied in situations when the
exact solution is unavailable. Second, and more important for this work is
that all the following constructions imply the computation of solutions for a
large set of different values of the spectral parameter $K^{2}$, while the
exact solutions involving the hypergeometric functions have been proved
considerably less practical than the approximate solutions obtained below. The
SPPS method allows one to construct a solution in the form of a power series
with respect to the parameter $K^{2}$ which is ideally suited for our
purposes. Compared to the use of exact solutions it gives us the possibility
to calculate a solution for different values of $K^{2}$ in a more efficient
way. From Figs.~(1) and (2) one can assess the excellent agreement displayed
by the zeroth-cell solutions obtained using both methods for a given set of
the spectral parameter $K^{2}$.

\subsection{Self-matching cell solutions}

Now we begin with the equation for $f_{c}$ on the zeroth cell $x\in
\lbrack0,\Lambda]$ in order to construct the so-called self-matching pair of
independent cell solutions $F_{\pm}$\cite{J49}. To obtain these special
solutions it is necessary first to have two linearly independent solutions
$f_{1}$ and $f_{2}$ satisfying the following initial conditions%
\begin{align}
f_{1}(0,K^{2}) &  =1,\qquad f_{2}(0,K^{2})=0,\label{y1y2 initial}\\
f_{1}^{\prime}(0,K^{2}) &  =0,\qquad f_{2}^{\prime}(0,K^{2})=1.\nonumber
\end{align}
The method of spectral parameter power series (SPPS) \cite{vvk,KP} gives these
solutions in explicit form as follows. Let $f_{0}$ be a particular solution of
$-f_{0}^{\prime\prime}+\nu_{1}f_{0}=K_{0}^{2}f_{0}$ such that $f_{0}\in
C^{2}(0,\Lambda)$ together with $\frac{1}{f_{0}}$ are bounded on $[0,\Lambda
]$. The general solution $f_{c}$ has the form $f_{c}=C_{1}f_{1}+C_{2}f_{2}$,
with
\begin{align}
f_{1}(x) &  =\frac{f_{0}(x)}{f_{0}(0)}\widetilde{\Sigma}_{0}(x)+f_{0}^{\prime
}(0)f_{0}(x)\Sigma_{1}(x),\nonumber\\
& \label{f1 f2}\\
f_{2}(x) &  =-f_{0}(0)f_{0}(x)\Sigma_{1}(x),\nonumber
\end{align}
where $\widetilde{\Sigma}_{0}$ and $\Sigma_{1}$ are the spectral parameter
power series $\widetilde{\Sigma}_{0}(x)=\sum_{\,n=0}^{\infty}\widetilde
{X}^{(2n)}(x)(K^{2}-K_{0}^{2})^{n}$, $\Sigma_{1}(x)=\sum_{n=1}^{\infty
}X^{(2n-1)}(x)(K^{2}-K_{0}^{2})^{n-1}$ with the coefficients $\widetilde{X}$,
$X$ given by the following recursive relations
\begin{equation}
\widetilde{X}^{(0)}\equiv1,\qquad X^{(0)}\equiv1,
\end{equation}

\bigskip%

\begin{equation}
\tilde{X}^{(n)}(x)=%
\begin{cases}
\int_{0}^{x}\tilde{X}^{(n-1)}(\xi)f_{0}^{2}(\xi)d\xi\qquad\mathrm{for}%
\,\mathrm{an}\,\mathrm{odd}\,n\\
\\
\int_{0}^{x}\tilde{X}^{(n-1)}(\xi)\frac{d\xi}{-f_{0}^{2}(\xi)}\qquad
\ \ \ \ \mathrm{for}\,\mathrm{an}\,\mathrm{even}\,n
\end{cases}
\label{K1}%
\end{equation}

\bigskip%

\begin{equation}
X^{(n)}(x)=%
\begin{cases}
\int_{0}^{x}X^{(n-1)}(\xi)\frac{d\xi}{-f_{0}^{2}(\xi)}\qquad
\ \ \ \ \ \mathrm{for}\,\mathrm{an}\,\mathrm{odd}\,n\\
\\
\int_{0}^{x}X^{(n-1)}(\xi)f_{0}^{2}(\xi)d\xi\qquad\ \mathrm{for}%
\,\mathrm{an}\,\mathrm{even}\,n~.
\end{cases}
\label{K2}%
\end{equation}
The solution $f_{0}$ is given by (\ref{f0}) and corresponds to the particular
value of $K_{0}^{2}=-\left(  k_{0}+s\right)  ^{2}$ which represents a band edge.

One can check by a straightforward calculation that the solutions $f_{1}$ and
$f_{2}$ fulfill the initial conditions (\ref{y1y2 initial}). In Fig. 1 we plot
the series solutions $f_{1}$ and $f_{2}$ evaluated by (\ref{f1 f2}) by a solid
line and the markers represent the exact solutions calculated taking the
appropriate linear combinations of $f_{c,1}$ and $f_{c,2}$ \ given
respectively by (\ref{2-fcfin1}) and (\ref{2-fcfin2}) in order to fulfill the
same initial conditions.%

\includegraphics[
height=2.6273in,
width=4.5965in
]%
{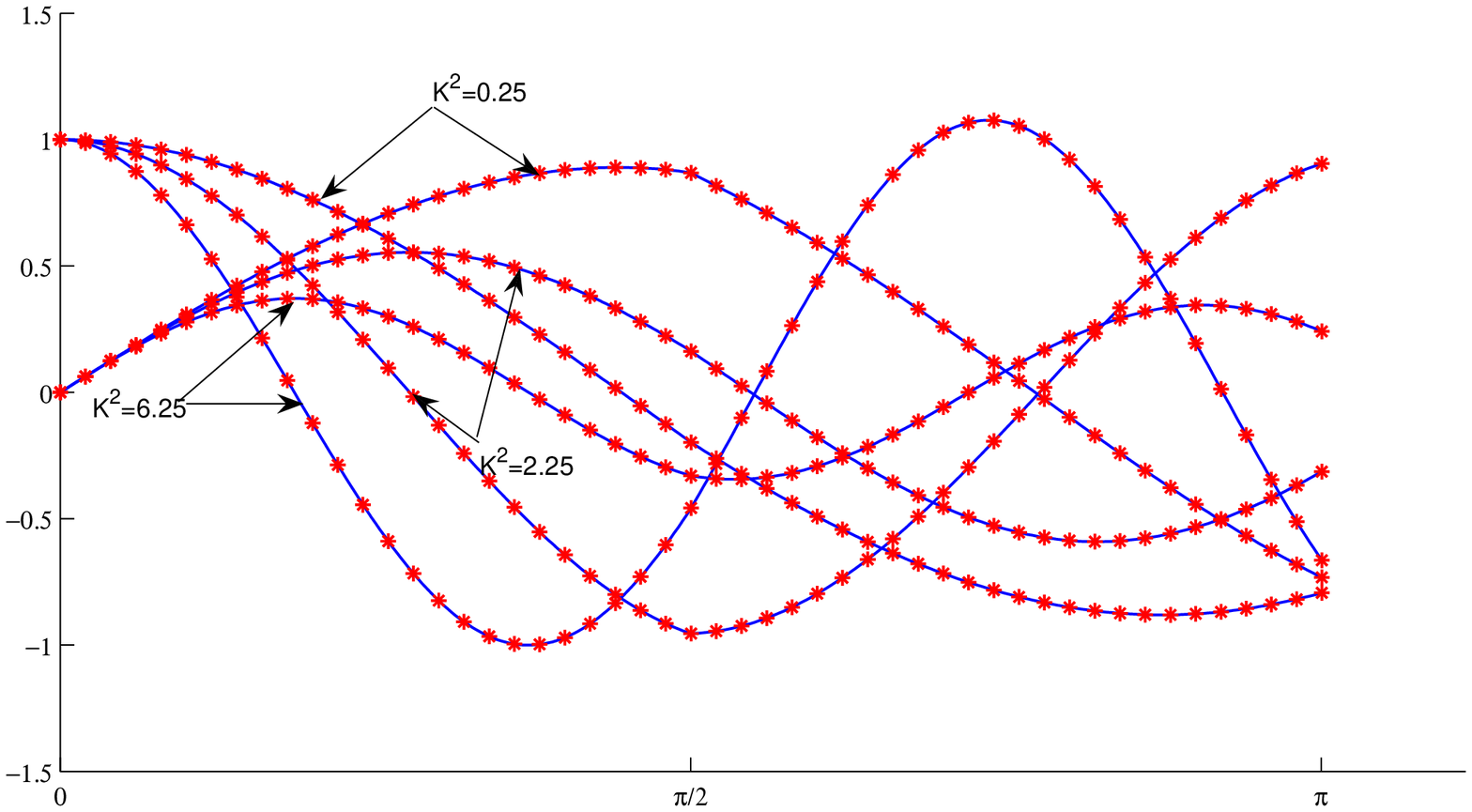}
\\
{\small Fig. 1. The solid lines represent the series solutions $f_{1}$ and
$f_{2}$ evaluated by (\ref{f1 f2}) and the markers represent the exact
solutions (\ref{2-fcfin1}) and (\ref{2-fcfin2}). The values of the parameters $S$ and $k_0$ are 0.1$i$ and 1, respectively.
}

\medskip

Denoting now
\begin{align*}
f_{1}(\Lambda,K^{2}) &  =a_{11}(K^{2}),\quad f_{2}(\Lambda,K^{2})=a_{12}%
(K^{2}),\\
f_{1}^{\prime}(\Lambda,K^{2}) &  =a_{21}(K^{2}),\quad f_{2}^{\prime}%
(\Lambda,K^{2})=a_{22}(K^{2}),
\end{align*}
one can write the self-matching solutions $F_{\pm}$ in the form%
\[
F_{\pm}(x,K^{2})=f_{1}(x,K^{2})+\alpha_{\pm}f_{2}(x,K^{2}),
\]
where $\alpha_{\pm}$ solve the equation $a_{12}\alpha_{\pm}^{2}+(a_{11}%
-a_{22})\alpha_{\pm}-a_{21}=0$ \cite{J49}.

To obtain the self-matching solutions to the equation (\ref{S-1comp}) we first
construct the solutions $g_{1}$ and $g_{2}$ which satisfy the initial
conditions $g_{1}(0,K^{2})=g_{2}^{\prime}(0,K^{2})=1$ and $g_{2}%
(0,K^{2})=g_{1}^{\prime}(0,K^{2})=0$. For this taking the following linear
combinations%
\[
\tilde{f}_{1}(x)=\frac{1}{K^{2}}\left(  Sf_{1}(x)+\left(  S^{2}+K^{2}\right)
f_{2}(x)\right)  \text{ and }\tilde{f}_{2}(x)=\frac{1}{K^{2}}\left(
f_{1}(x)+Sf_{2}(x)\right)
\]
and applying the Darboux transformation (\ref{trans g}) to them gives%
\begin{equation}
g_{1}(x)=\tilde{f}_{1}^{\prime}(x)-\Phi_{1}(x)\tilde{f}_{1}(x)\text{ and
}g_{2}(x)=\tilde{f}_{2}^{\prime}(x)-\Phi_{1}(x)\tilde{f}_{2}(x).\label{g1g2}%
\end{equation}
Illustrative plots of the latter solutions are displayed in Fig.~(2) in solid
lines, while the markers correspond to the exact formulas (\ref{g1}) and
(\ref{g2}). Notice also that all the singularities of the above solutions are
contained in the Darboux transformation function $\Phi_{1}(x)=S-k_{0}\tan
k_{0}x$.%

\includegraphics[
height=2.559in,
width=4.6103in
]%
{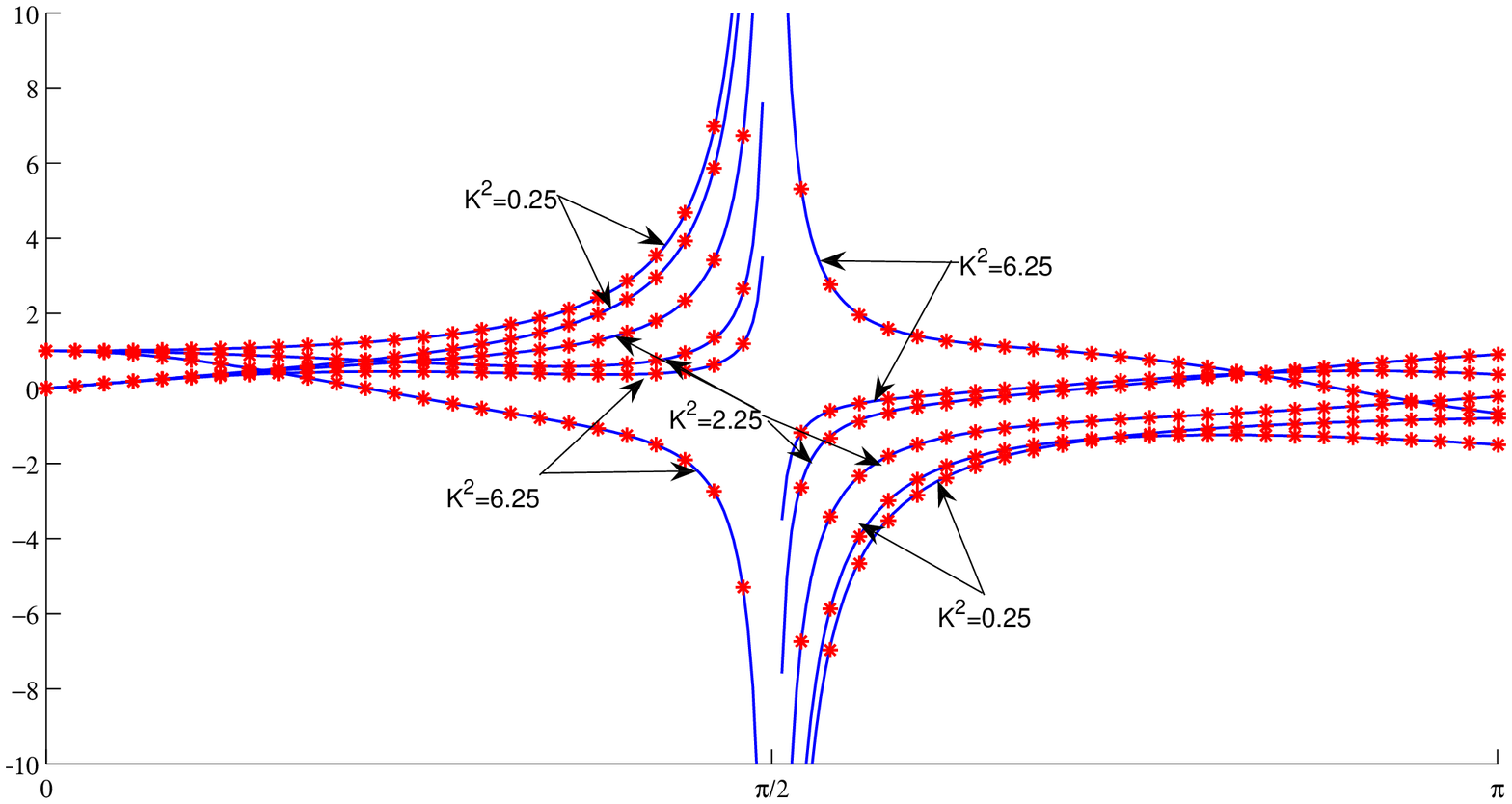}
\\
{\small Fig. 2. Series and exact solutions $g_{1}$ and
$g_{2}$ evaluated by (\ref{g1g2}) and
(\ref{g1}) and (\ref{g2}), respectively, for the same parameters as in Fig.~1.}

\medskip By analogy with the $f$-case we denote
\begin{align*}
g_{1}(\Lambda,K^{2}) &  =b_{11}(K^{2}),\quad g_{2}(\Lambda,K^{2})=b_{12}%
(K^{2}),\\
g_{1}^{\prime}(\Lambda,K^{2}) &  =b_{21}(K^{2}),\quad g_{2}^{\prime}%
(\Lambda,K^{2})=b_{22}(K^{2}),
\end{align*}
and the self-matching solutions $G_{\pm}$ have the form%
\[
G_{\pm}(x,K^{2})=g_{1}(x,K^{2})+\beta_{\pm}g_{2}(x,K^{2}),
\]
where $\beta_{\pm}$ are roots of the equation $b_{12}\beta_{\pm}^{2}%
+(b_{11}-b_{22})\beta_{\pm}-b_{21}=0$.

\subsection{Bloch solutions}

We are now in a position to write down the Bloch (quasi-periodic) solutions to
the equation (\ref{S-2comp}) through the whole range of $x$ divided as follows
$n\Lambda\leq x\prec(n+1)\Lambda$ for $n=0,\pm 1,\pm 2,\cdots$%
\begin{equation}
f_{\pm}(x,K^{2})=r_{\pm}^{n}F_{\pm}(x-n\Lambda,K^{2})=r_{\pm}^{n}\left[
f_{1}(x-n\Lambda,K^{2})+\alpha_{\pm}f_{2}(x-n\Lambda,K^{2})\right]
.\label{f+-}%
\end{equation}
The Bloch factors $r_{\pm}$ are a measure of the rate of increase (or
decrease) in magnitude of the self-matching solutions $F_{\pm}(x,K^{2})$ when
one goes from the left end of the cell to the right one, i.e.,
\[
r_{\pm}(K^{2})=\frac{F_{\pm}(\Lambda,K^{2})}{F_{\pm}(0,K^{2})}~.
\]
The values of $r_{\pm}$ can be also written as
\[
r_{\pm}(K^{2})=\frac{1}{2}\left(  D_{f}(K^{2})\mp\sqrt{D_{f}^{2}(K^{2}%
)-4}\right)  ~,
\]
where $D_{f}(K^{2})$ denotes Hill's discriminant (also known as Lyapunov
function) associated with (\ref{S-2comp}) \cite{Magnus} and $D_{f}%
(K^{2})=a_{11}+a_{22}$. For equation (\ref{S-1comp}) the Hill discriminant is
given by $D_{g}(K^{2})=b_{11}+b_{22}$. Using the relations (\ref{g1g2})
between the solutions $g_{1}$, $g_{2}$ and $f_{1}$, $f_{2}$ the identity
$D_{f}(K^{2})\equiv D_{g}(K^{2})$ can be easily obtained. This means that the
Bloch factors for the quasi-periodic solutions to equation (\ref{S-1comp}) are
the same as for (\ref{S-2comp}). Thus, for a numerable set of cells,
$n\Lambda\leq x\prec(n+1)\Lambda$ for $n=0,1,2,\cdots$, one can write
\begin{equation}
g_{\pm}(x,K^{2})=r_{\pm}^{n}G_{\pm}(x-n\Lambda,K^{2})=r_{\pm}^{n}\left[
g_{1}(x-n\Lambda,K^{2})+\beta_{\pm}g_{2}(x-n\Lambda,K^{2})\right]
.\label{g+-}\\
\end{equation}

In Fig.~3 the Bloch solutions (\ref{f+-}) and (\ref{g+-}) are plotted with
solid and dotted lines respectively.%

\includegraphics[
height=2.5097in,
width=4.9995in
]%
{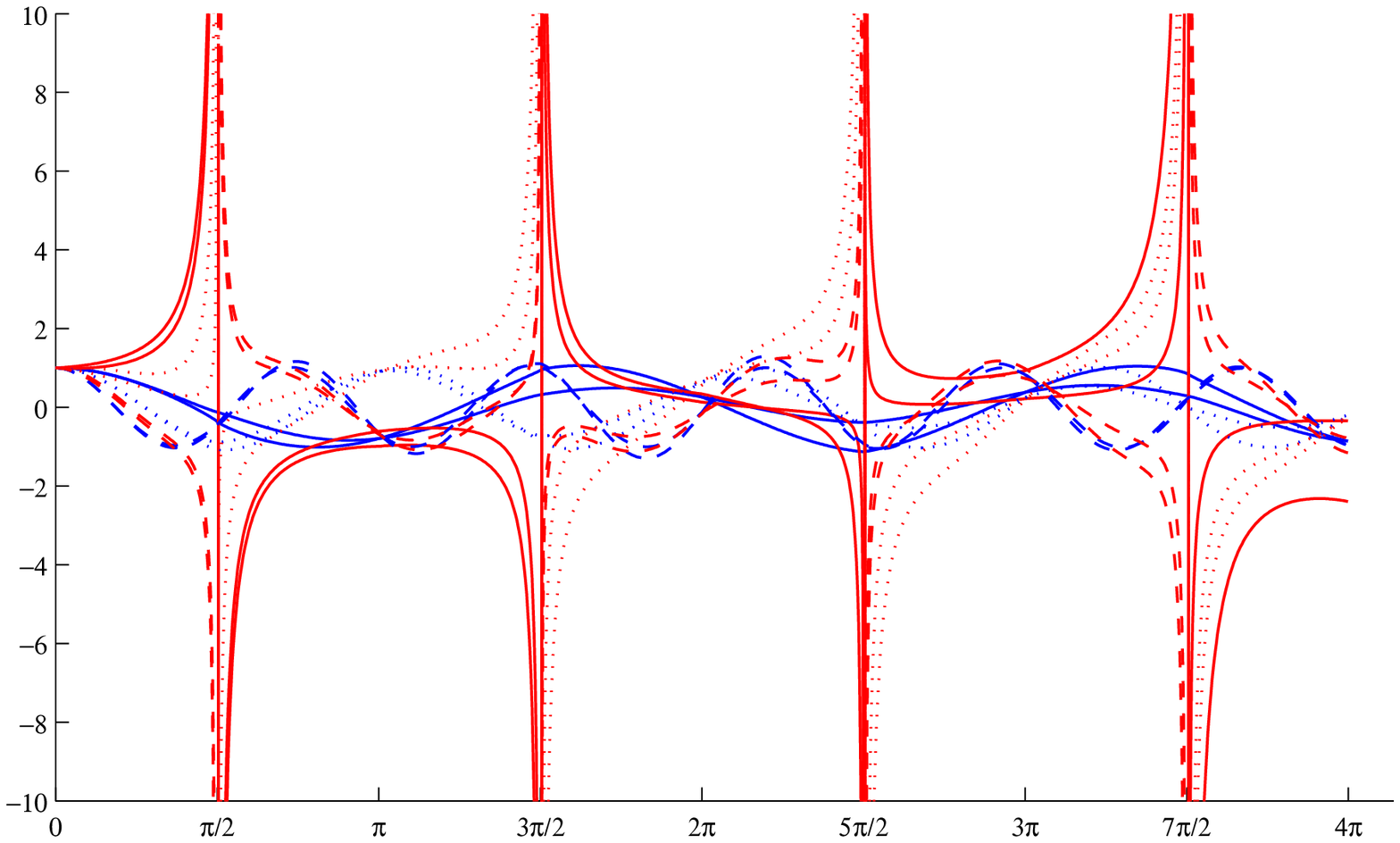}
\\
{\small Fig.~3. The blue lines are the solutions $f_{\pm}$
given by (\ref{f+-}) and the red lines represent the solutions
$g_{\pm}$ given by (\ref{g+-}) for the same values of the parameters as previously and for
$K^2=0.25$ (solid
lines), $2.25$ (dotted lines), and $6.25$ (dashed lines).}

\subsection{A power series representation for the Hill discriminant}

The Hill discriminant (Lyapunov function) allows one to describe the spectrum
of periodic differential equations. Namely, the spectrum of (\ref{S-2comp})
and (\ref{S-1comp}) is given by the following set \cite{Eastham} $\left\{
K^{2}:D_{f}(K^{2})\in\mathbf{R}\text{ and }\left\vert D_{f}(K^{2})\right\vert
\leq2\right\}  $. The expression for $D_{f}(K^{2})=a_{11}+a_{22}$ can be
written in a simple explicit form. For this we write $a_{11}$ and $a_{22}$ in
form of a spectral parameter power series using (\ref{f1 f2}) and taking into
account that $\frac{d}{dx}\Sigma_{1}(x)=\frac{1}{-f_{0}^{2}(x)}\Sigma_{0}(x)$,
where $\Sigma_{0}(x)=\sum_{n=0}^{\infty}X^{(2n)}(x)(K^{2}-K_{0}^{2})^{n}$:%
\[
a_{11}=\frac{f_{0}(\Lambda)}{f_{0}(0)}\widetilde{\Sigma}_{0}(\Lambda
)+f_{0}^{\prime}(0)f_{0}(\Lambda)\Sigma_{1}(\Lambda)\text{ and }a_{22}
=-f_{0}(0)f_{0}^{\prime}(\Lambda)\Sigma_{1}(\Lambda)+\frac{f_{0}(0)f_{0}
(\Lambda)}{f_{0}^{2}(\Lambda)}\Sigma_{0}(\Lambda)~.
\]
Since $f_{0}(x)$ is a $\Lambda$-periodic function: $f_{0}(0)=f_{0}(\Lambda) $.
Finally, writing the explicit expressions for $\widetilde{\Sigma}_{0}%
(\Lambda)$ and $\Sigma_{0}(\Lambda)$ we obtain a representation for the
Hill discriminant associated with (\ref{S-2comp}) and (\ref{S-1comp})
\begin{equation}
D_{f}(K^{2})\equiv D_{g}(K^{2})=\sum_{n=0}^{\infty}\left(\tilde{X}
^{(2n)}(\Lambda)+X^{(2n)}(\Lambda)\right)  (K^{2}-K_{0}^{2})^{n}
\text{.}\label{D}%
\end{equation}
Thus, only one particular nodeless periodic solution $f_{0}(x)$ of (\ref{S-2comp}) is
needed for the construction of the Hill discriminant $D_{f}(K^{2})$.
There are other known series representations of the Hill discriminant, see \cite{Magnus,Jagerman}.
Nevertheless none of them allows one to represent it as a spectral
parameter power series which is extremely useful for calculations involving different
values of the spectral parameter.

Figure~4 shows the plot of $D_{f}(K^{2})$ which we evaluate in two ways. With
the solid line we plot the function $D_{f}(K^{2})$ obtained by means of the
SPPS solutions given by (\ref{D}) and the markers correspond to $D_{f}(K^{2})$
obtained with the exact solutions (\ref{2-fcfin1}) and (\ref{2-fcfin2}). From
(\ref{D}) the advantage of the SPPS method for calculating the Hill discriminant
and hence the corresponding spectrum can be assessed. Using the SPPS
the calculation of the value of the Hill discriminant for every value of its
argument reduces to a simple substitution of the value of $K^{2}$ into an
easily evaluated expression (\ref{D}). The values of $K^{2}$ for which
$D_{f}(K^{2})=\pm2$ correspond to the band edges of the spectrum. Notice that
$D_{f}(K^{2})=\pm2$ is in accordance with (\ref{per}) and (\ref{antiper}),
namely $D_{f}(K^{2})=2$ exactly for $K^{2}=(k_{0}-iS)^{2}-(2nk_{0})^{2}$ and
$D_{f}(K^{2})=-2$ when $K^{2}=(k_{0}-iS)^{2}-((2n+1)k_{0})^{2}$.

When $D(K^{2})\neq\pm 2$, the general solutions of (\ref{S-2comp}) and
(\ref{S-1comp}) have the form (understanding that it refers to the whole $x$
axis henceforth)
\[
f_{c}(x)=C_{+}f_{+}(x)+C_{-}f_{-}(x)\quad\text{and\quad}g_{c}(x)=\tilde{C}%
_{+}g_{+}(x)+\tilde{C}_{-}g_{-}(x)
\]

For the values of $K^{2}$ giving $D_{f}(K^{2})=$ $\pm2$, two pairs of
independent solutions $f_{+}(x,K^{2})$, $f_{-}(x,K^{2})$ and $g_{+}(x,K^{2})$,
$g_{-}(x,K^{2})$\ reduce to a pair of a single solutions $f_{+}(x,K^{2})\equiv
f_{-}(x,K^{2})$ and $g_{+}(x,K^{2})\equiv g_{-}(x,K^{2}) $ , but there is a
definite prescription for the construction of an independent second solution
\cite{J49}.%

\includegraphics[
height=2.6273in,
width=5.6464in
]%
{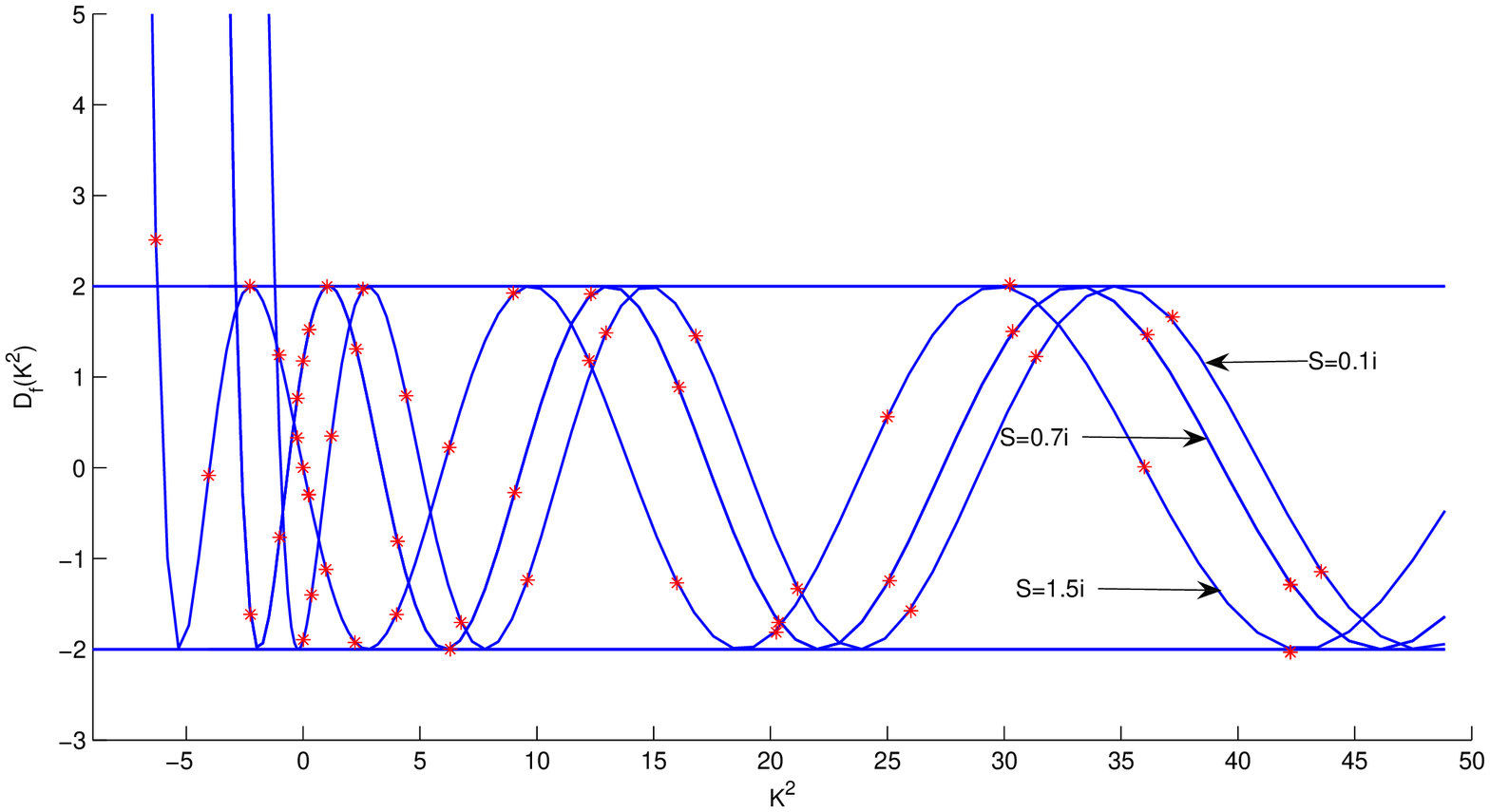}
\\
{\small Fig.~4. The Hill discriminant $D_{f}(K^{2})$ evaluated using SPPS solutions
(\ref{f1 f2}) (solid line) and by the exact solutions (\ref{2-fcfin1}) and
(\ref{2-fcfin2}) ( markers) for $S=\{0.1i,0.7i,1.5i\}$.}%

\section{Conclusions}

\medskip

\noindent We have used one particular solution of simple Riccati equations of
constant parameters to build the corresponding supersymmetric partner
Sturm-Liouville equations. The latter equations are solved analytically in
terms of hypergeometric functions. Furthermore we worked with Kravchenko's
spectral parameter power series solutions that are better suited from the
algorithmic (numerical) standpoint and allows an easy implementation of the
old self-matching procedure of H.M. James \cite{J49} for solving periodic
Sturm-Liouville problems of Schrödinger type in terms of Bloch solutions. We
also obtain an effective power series representation of the Hill discriminant
in terms of the Kravchenko series. The mathematical procedure expounded in
this paper can be applied to more general periodic SL equations that abound in
the area of nanostructured materials and in the form of periodic Helmholtz
equations in photonics. Other applications can be foreseen in the areas of
chirp technology, see for example Refs. \cite{Biyarin} and \cite{Genty}.

\bigskip

\noindent
The first author would like to thank CONACyT for a postdoctoral fellowship allowing her to
work in IPICyT.

\end{document}